\documentclass[conference]{IEEEtran}
\IEEEoverridecommandlockouts

\usepackage{cite}
\usepackage{amsmath,amssymb,amsfonts}
\usepackage{algorithmic}
\usepackage{graphicx}
\usepackage[table]{xcolor}
\usepackage{textcomp}
\usepackage{xcolor}
\usepackage{booktabs} 
\usepackage{listings}
\usepackage{orcidlink}

\def\BibTeX{{\rm B\kern-.05em{\sc i\kern-.025em b}\kern-.08em
    T\kern-.1667em\lower.7ex\hbox{E}\kern-.125emX}}
\begin{document}

\title{Explainable Galaxy Interaction Prediction with Hybrid Attention Mechanisms
}
\author{
    \IEEEauthorblockN{
        Sathwik Narkedimilli \orcidlink{0009-0001-9019-6283}\IEEEauthorrefmark{2},
        Satvik Raghav\IEEEauthorrefmark{3}, 
        Om Mishra\IEEEauthorrefmark{1},
        Mohan Kumar\IEEEauthorrefmark{6},
        Aswath Babu H\IEEEauthorrefmark{4},\\
        Tereza Jerabkova\IEEEauthorrefmark{7},
        Manish M\IEEEauthorrefmark{5}, and
        Sai Prashanth Mallellu\IEEEauthorrefmark{8}
    }
    \vspace{0.5em} 
    \IEEEauthorblockA{
        \IEEEauthorrefmark{2}Department of Electrical and Computer Engineering, National University of Singapore (NUS), Singapore\\
    }
    \IEEEauthorblockA{
        \IEEEauthorrefmark{3}School of Computing Technologies, RMIT University, Melbourne, VIC 3000, Australia\\
    }
    \IEEEauthorblockA{
        \IEEEauthorrefmark{1}Department of Electronics and Communication Engineering, Indian Institute of Information Technology Dharwad, India\\
    }
    \IEEEauthorblockA{
        \IEEEauthorrefmark{6}Senior Data Engineer at SDI, Indian Air Force (IAF), India\\
    }
    \IEEEauthorblockA{
        \IEEEauthorrefmark{4}Department of Arts, Science, and Design, Indian Institute of Information Technology Dharwad, India\\
    }
    \IEEEauthorblockA{
        \IEEEauthorrefmark{7}Department of Theoretical Physics and Astrophysics, Faculty of Science, Masaryk University, Czech Republic
    }
    \IEEEauthorblockA{
        \IEEEauthorrefmark{5} NVIDIA, India
    }
    \IEEEauthorblockA{
        \IEEEauthorrefmark{8}Department of Computer Science \& Engineering, Symbiosis Institute of Technology, Hyderabad Campus,\\
        Symbiosis International University, Pune, India\\
    }
    \vspace{0.5em}
    \IEEEauthorblockA{
    Emails: 
    sathwik.narkedimilli@ieee.org; satvikraghav007@gmail.com; 23bec035@iiitdwd.ac.in; mk.elitetech@gmail.com;\\ aswath@iiitdwd.ac.in; tereza.jerabkova@eso.org; mmodani@nvidia.com; saiprashanth08@ieee.org
  }
}

\maketitle

\begin{abstract}
Galaxy interaction classification remains challenging due to complex morphological patterns and the limited interpretability of deep learning models. We propose an attentive neural ensemble that combines \textbf{AG-XCaps}, \textbf{H-SNN}, and \textbf{ResNet-GRU} architectures, trained on the \textbf{Galaxy Zoo DESI} dataset and enhanced with \textbf{LIME} to enable explainable predictions. The model achieves \textbf{Precision = 0.95}, \textbf{Recall = 1.00}, \textbf{F1 = 0.97}, and \textbf{Accuracy = 96\%}, outperforming a Random Forest baseline by significantly reducing false positives (\textbf{23 vs. 70}). This lightweight (\textbf{0.45 MB}) and scalable framework provides an interpretable and efficient solution for large-scale surveys such as \textbf{Euclid} and \textbf{LSST}, advancing data-driven studies of galaxy evolution.
\end{abstract}


\section{Introduction \& Related Works}

Galaxies, the fundamental units of the universe, display diverse morphologies that reflect their complex formation and evolutionary histories shaped by gravitational interactions over billions of years. Galaxy interactions, encompassing mergers, tidal distortions, and minor encounters, drive structural transformations such as tidal tails, bridges, and rings, while often inducing intense star formation that exposes underlying gravitational and hydrodynamic mechanisms \cite{toomre1972galactic, narkedimilli2025predicting}. Observed across redshifts, these processes illuminate transitions between morphological types, such as spirals evolving into ellipticals, and the influence of environmental conditions on galactic evolution \cite{pearson2019identifying}. Large-scale surveys like the DESI Legacy Imaging Surveys and Galaxy Zoo have cataloged millions of galaxies with detailed morphological classifications \cite{walmsley2023galaxy}. However, traditional analytical methods remain limited in their ability to predict interaction outcomes or to link them to specific physical drivers, leaving key astrophysical connections unexplored.

To address this gap, our research aims to enhance predictive accuracy while maintaining astrophysically interpretable models for galaxy interactions. With upcoming missions such as Euclid and LSST expected to produce vast datasets, the need for automated yet transparent modeling tools is increasingly critical. Although machine learning models achieve high accuracy in static morphology classification, their black-box nature limits scientific insight by obscuring causal reasoning \cite{cao2024galaxy}. Our study seeks to overcome this limitation by forecasting interaction outcomes, such as merger likelihood and morphological evolution, while uncovering the governing physical processes, thereby improving trust and facilitating the integration of explainable models into observational pipelines and cosmological simulations \cite{nelson2019illustristng}.
Our research introduces the following key contributions to the field of galaxy interaction prediction:

Galaxy morphology studies have advanced significantly through machine learning and cosmological analyses, offering new perspectives on galactic evolution. Cao \textit{et al.} \cite{cao2024galaxy} proposed a Convolutional Vision Transformer (CvT) that combines CNNs and transformers for large-scale morphological classification, achieving superior accuracy yet limited interpretability. Haslbauer \textit{et al.} \cite{haslbauer2022high} analyzed thin disk galaxy prevalence within the $\Lambda$CDM framework, finding observed thin disks more frequent than simulations predict, exposing modeling limitations in merger and feedback processes. Laishram \textit{et al.} \cite{laishram2024insights} studied galaxies at \(z \sim 1.5\) using [O II] emitters, linking filamentary environments to disturbed morphologies and enhanced star formation, offering critical observational insights into interaction-driven evolution.  

Deep learning and manifold learning approaches continue to enrich research on galaxy morphology. Urechiatu \textit{et al.} \cite{urechiatu2024improved} improved CNN-based morphology classification, achieving high accuracy across spirals, ellipticals, and irregulars, yet lacking explainability. Semenov \textit{et al.} \cite{tsizh5057237galaxy} employed manifold learning for unsupervised feature extraction, efficiently distinguishing morphological types while improving interpretability through latent feature visualization. Despite these advances, most existing models prioritize classification performance over causal transparency or predictive modeling of galaxy interactions. These works collectively motivate our study’s focus on explainable neural ensembles that merge predictive accuracy with astrophysical interpretability in galaxy interaction prediction.

The reviewed literature highlights significant progress in galaxy morphology studies, where deep learning models by Cao et al.~\cite{cao2024galaxy} and Urechiatu~\cite{urechiatu2024improved} achieve high classification accuracy. Semenov et al.~\cite{tsizh5057237galaxy} enhance interpretability through manifold learning. Observational works by Haslbauer et al.~\cite{haslbauer2022high} and Laishram et al.~\cite{laishram2024insights} emphasize the roles of mergers and environmental effects. Yet, deep learning approaches remain opaque, manifold methods lack predictive interaction modeling, and observational studies are not integrated with machine learning. Our research bridges these gaps by introducing an explainable, interaction-focused model using attentive neural ensembles to unify predictive power with astrophysical interpretability. The key contributions include:

\begin{itemize}
    \item Hybrid AG-XCaps, H-SNN, and ResNet-GRU model for galaxy interaction features.  
    \item LIME-based interpretation of key morphological drivers.  
    \item \textbf{Astronomy Impact:} Redshift-aware morphology and merger analysis for Euclid and LSST.  
\end{itemize}

\section{System Model}

\subsection{Dataset}

The dataset proposed by Walmsley et al.~\cite{walmsley2023galaxy} contains morphology measurements for 8.67 million galaxies within the footprint of the DESI Legacy Imaging Surveys. It includes 41 columns such as unique identifiers (\texttt{dr8Id}), celestial coordinates (\texttt{RAdeg}, \texttt{DEdeg}), and object indices (\texttt{brickid}, \texttt{objid}). The remaining attributes represent predicted vote fractions for various morphological characteristics: smooth or featured (\texttt{SFSM}, \texttt{SFFDF}, \texttt{SFAF}); disk edge-on (\texttt{DEOYes}, \texttt{DEONo}); spiral arms (\texttt{SAYes}, \texttt{SANo}); bar type (\texttt{BS}, \texttt{BW}, \texttt{BNo}); bulge size (\texttt{BSD}, \texttt{BSL}, \texttt{BSM}, \texttt{BSS}, \texttt{BSNo}); roundness (\texttt{RR}, \texttt{RIB}, \texttt{RCS}); edge-on bulge shape (\texttt{EOBB}, \texttt{EOBNo}, \texttt{EOBR}); spiral winding (\texttt{SWT}, \texttt{SWM}, \texttt{SWL}); spiral arm count (\texttt{SAC1}, \texttt{SAC2}, \texttt{SAC3}, \texttt{SAC4}, \texttt{SAC4+}, \texttt{SACCT}); and handling missing values (replacing them with zeros) (\texttt{MMiD}, \texttt{MMaD}, \texttt{MM}). Metadata confirms 41 distinct morphological features recorded across more than 8.6 million galaxy entries.

The catalog was created by training deep learning models on Galaxy Zoo volunteer responses, leveraging newly collected votes for DESI-LS DR8 images and historical votes from previous Galaxy Zoo projects. These models predict the number of volunteers who would select each answer to the morphological questions, automating and standardizing the extraction of detailed galaxy features. Crucially, since these labels are derived from crowd-sourced votes, they represent probabilistic classifications that may contain human bias or disagreement, rather than absolute physical ground truths. The methodology enables extensive sky coverage and provides valuable insights into galaxy morphology by correlating automated predictions with observed features from the DESI Legacy Imaging Surveys.

\subsection{Data Pre-Processing and Feature Engineering}

Data preprocessing begins by handling missing values (replacing them with zeros), removing duplicates, converting data types, and removing outliers. The target variable \texttt{InteractionActivity} is derived by comparing the \texttt{MNo} feature to the maximum of \texttt{MMiD}, \texttt{MMaD}, and \texttt{MM}—with the result converted into a categorical format—and the dataset is further preprocessed by creating a new target variable \texttt{InteractionActivity} based on the following mathematical assumption:

\fbox{\resizebox{0.8\linewidth}{!}{$
\text{InteractionActivity} =
\begin{cases}
0, & \text{if } \text{MNo} > \max\{\text{MMiD},\, \text{MMaD},\, \text{MM}\} \\
1, & \text{otherwise.}
\end{cases}
$}}

\vspace{2pt}

In this step, the variable \texttt{InteractionActivity} is defined by comparing the \texttt{MNo} feature against the maximum value of the features \texttt{MMiD}, \texttt{MMaD}, and \texttt{MM}. These features are described as follows:
In the dataset, four columns represent the likelihood of a galaxy being involved in a galaxy interaction: MNo (Galaxy Interaction None Fraction), which is the fraction of volunteers who classified the galaxy as not undergoing any galaxy interaction; MMiD (Galaxy Interaction Minor Disturbance Fraction), representing those who identified a minor disturbance indicative of an early-stage galaxy interaction; MMaD (Galaxy Interaction Major Disturbance Fraction), representing those who classified the galaxy as undergoing a major disturbance with significant structural changes due to an ongoing interaction; and MM (Galaxy Interaction Fraction), indicating the fraction of volunteers who explicitly identified the galaxy as involved in a galaxy interaction, suggesting a late-stage interaction.

The reasoning behind the assumption is that if the probability of no interaction (MNo) exceeds the maximum likelihood among the indicators for disturbances or interactions (MMiD, MMaD, MM), the galaxy is classified as 0 (no galaxy interaction); otherwise, it is classified as 1 (galaxy interaction present). This approach simplifies galaxy categorization based on volunteer assessments and facilitates subsequent analyses using the dataset's other features.

The dataset is split into training and test sets in an 80–20 ratio, and the input variables are standardized using the StandardScaler. Feature engineering enhances data interpretability and model readiness through a structured workflow. Key predictors are first identified using Linear Discriminant Analysis (LDA) or Random Forest feature importance plots. Principal Component Analysis (PCA) is then applied to reduce dimensionality to 29 components, retaining maximum variance while minimizing redundancy. The optimized feature set is subsequently used to develop and evaluate predictive models for galaxy-interaction classification, achieving improved accuracy and computational efficiency.

\begin{figure}
    \centering
    \includegraphics[width=0.5\linewidth]{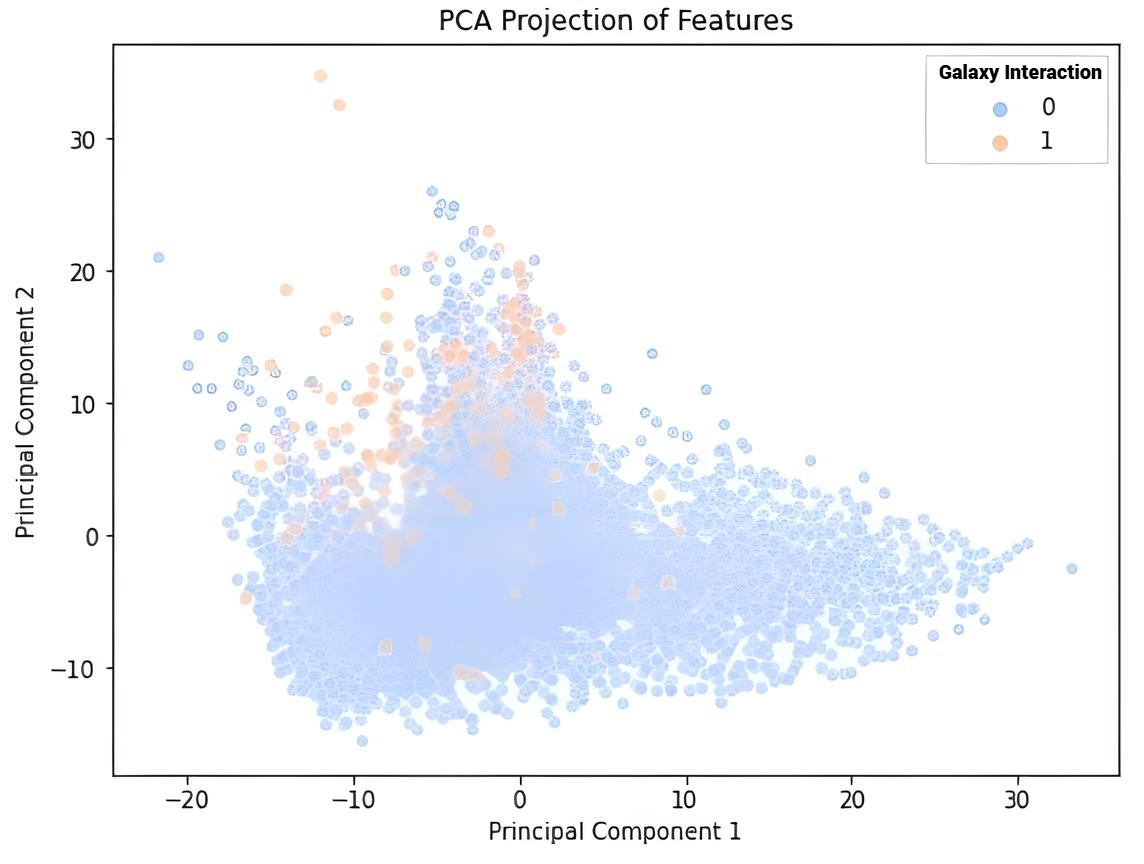}
    \caption{PCA visualization of the dataset}
    \label{pca}
\end{figure}

\begin{figure}
    \centering
    \includegraphics[width=0.5\linewidth]{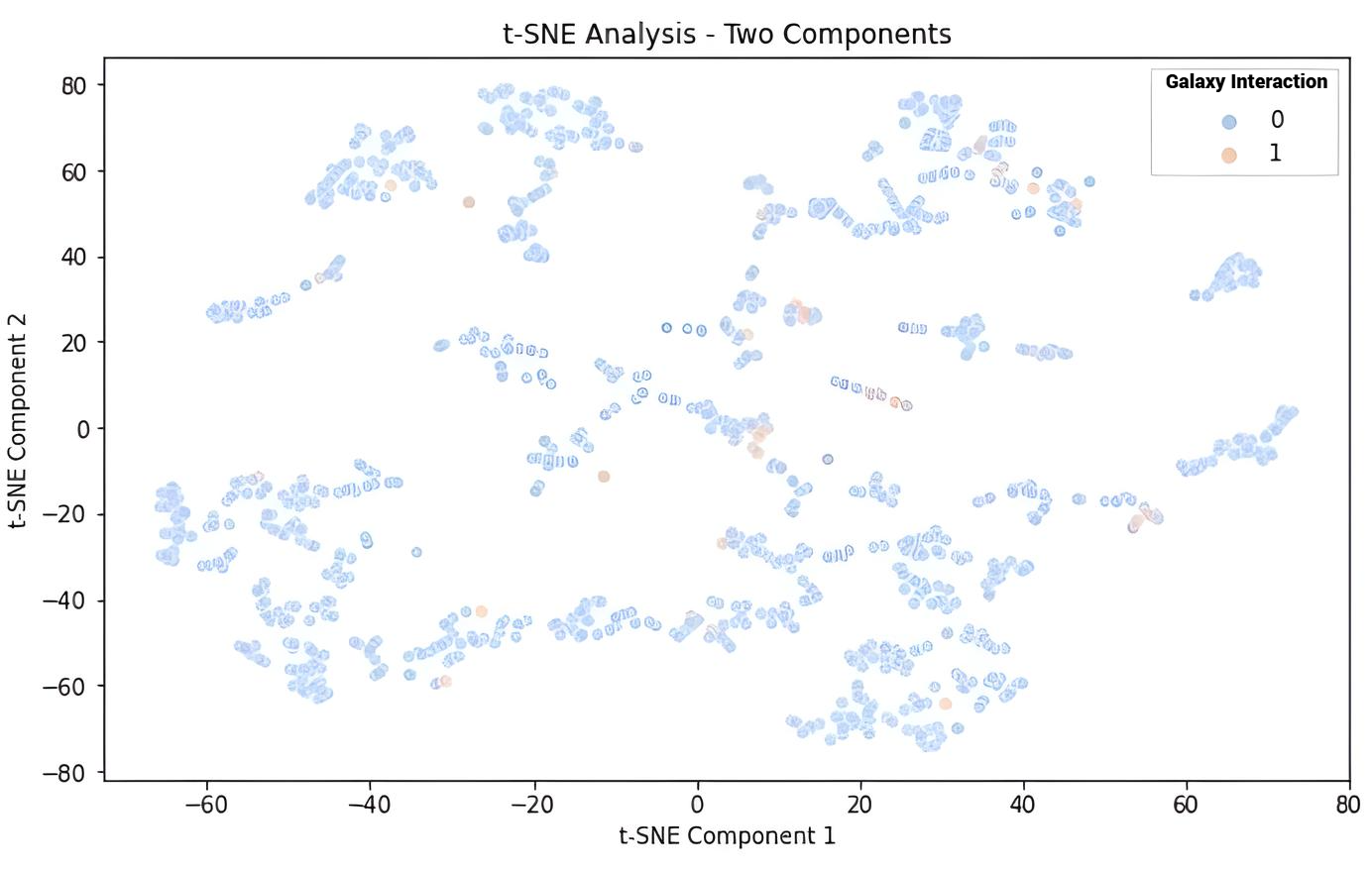}
    \caption{t-SNE visualization of the dataset}
    \label{tsne}
\end{figure}

From Fig.~\ref{pca} and Fig.~\ref{tsne}, it is evident that most galaxies (label 0) occupy a broad region, while those labeled as interactions (label 1, indicating detection of galaxy interaction) are relatively sparse. The PCA plot (Fig. 1) captures the most significant variance in the first two principal components, providing a linear overview of class separation. In contrast, the t-SNE plot (Fig. 2) provides a more nuanced, nonlinear perspective on local structures. This distribution mathematically demonstrates that galaxy interactions are rare.

\subsection{Workflow of the Proposed Model}

\begin{figure}
    \centering
    \includegraphics[width=0.95\linewidth]{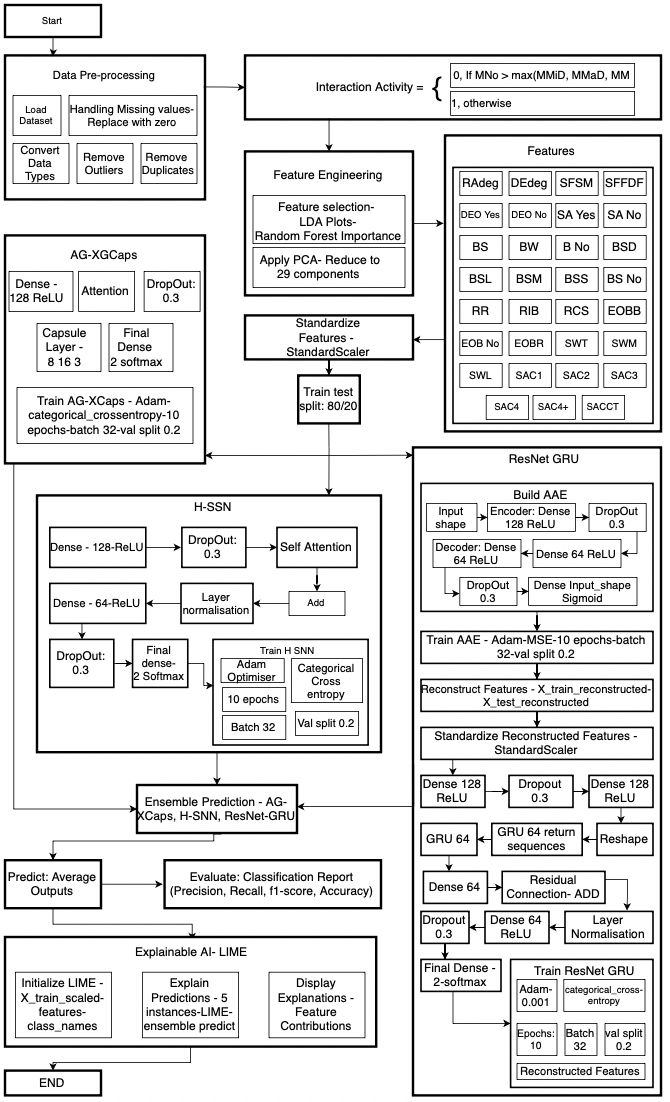}
    \caption{Workflow diagram of the Proposed Model}
    \label{workflowdiagram}
\end{figure}

\begin{figure}
    \centering
    \includegraphics[width=0.5\linewidth]{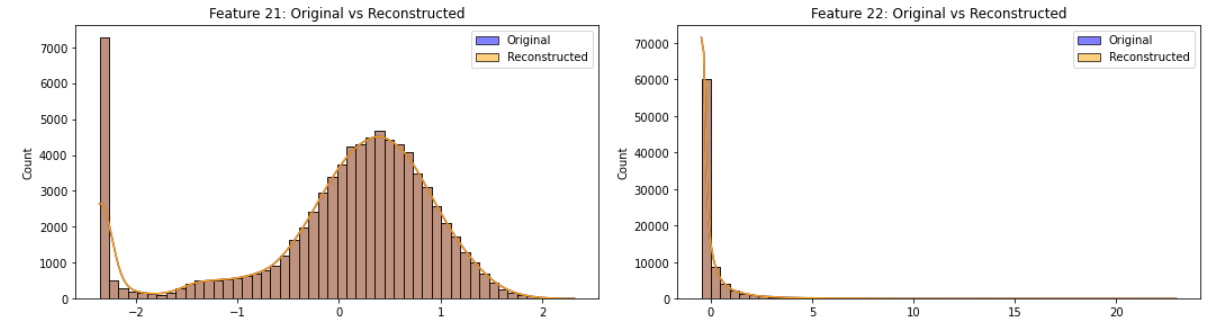}
    \caption{Example Transformation of Features via an AAE Before Input into the ResNet with GRU Layers model}
    \label{aaetransform}
\end{figure}

The workflow in Fig.~\ref{workflowdiagram} integrates three neural architectures: AG-XCaps, H-SNN, and ResNet-GRU, to predict comprehensive galaxy interactions. AG-XCaps includes a Dense-128-ReLU layer, an Attention block, and an 8-capsule (16D, three routings) layer, followed by a Dense-2-softmax output. H-SNN combines Dense-128-ReLU with 0.3 dropout, self-attention, Layer Normalization, Dense-64-ReLU, and final Dense-2-softmax. ResNet-GRU uses two Dense-128-ReLU layers (the first with a 0.3 dropout rate), two GRU-64 layers with a Dense-64 residual path, Layer Normalization, Dense-64-ReLU, and a Dense-2-softmax output. Trained on Autoencoder-transformed data (Fig.~\ref{aaetransform}), ResNet-GRU benefits from denoised, compressed morphological representations.

All models are trained for 10 epochs using Adam with categorical cross-entropy, a batch size of 32, and a 20\% validation split, except for AAE (Adam, MSE) and ResNet-GRU (Adam, learning rate 0.001). Each architecture addresses distinct aspects of galaxy interaction analysis: AG-XCaps captures hierarchical spatial features via capsules, H-SNN employs self-attention to emphasize key morphological traits, and ResNet-GRU models sequential or higher-order patterns from AAE-transformed data using residual and gated mechanisms. They provide a complementary framework well-suited for large-scale astronomical surveys with diverse and complex galaxy morphologies.

After individual training, the models’ outputs are averaged to obtain an ensemble prediction for the \texttt{InteractionActivity} class, thereby leveraging each architecture's diversity to improve overall accuracy and robustness. LIME (Local Interpretable Model-Agnostic Explanations) is then applied to a subset of test instances, providing transparency into how each feature influences the ensemble’s predictions. By highlighting the contributions of features such as tidal distortions, LIME ties the model outputs to astrophysical phenomena, ensuring that the system remains predictive and interpretable, an essential requirement for large-scale astronomical surveys such as Euclid and LSST.

\begin{figure}
    \centering
    \includegraphics[width=0.5\linewidth]{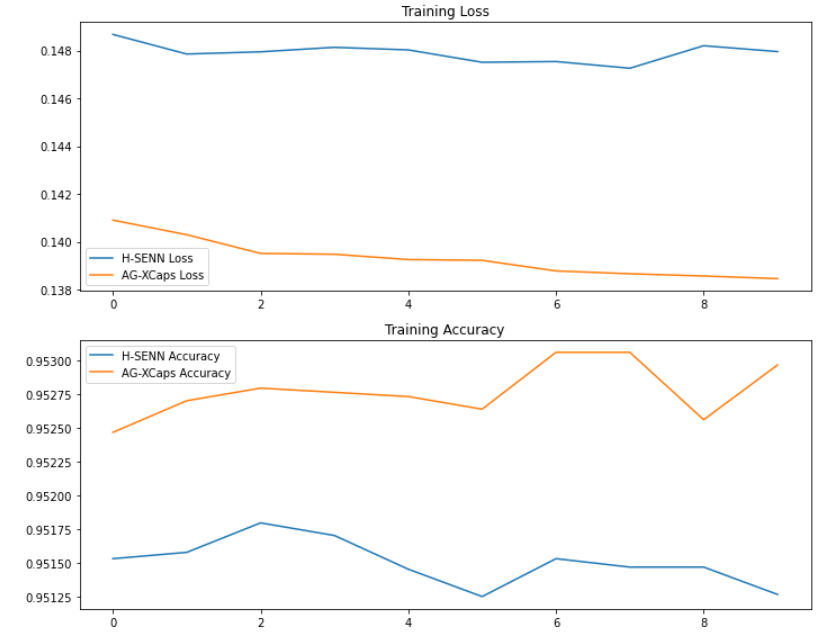}
    \caption{Training and validation loss curves for the H-SENN and AG-XCaps Models}
    \label{2algotesttrainloss}
\end{figure}

\begin{figure}
    \centering
    \includegraphics[width=0.5\linewidth]{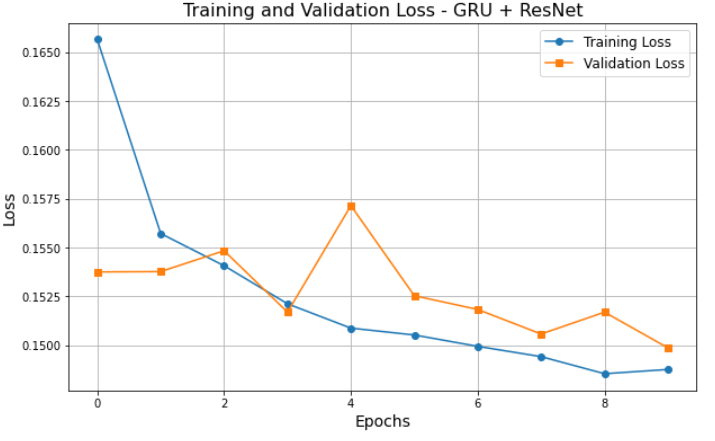}
    \caption{Training and validation loss curves for the ResNet-GRU model}
    \label{onealgotesttrainloss}
\end{figure}

We compute standard performance metrics to evaluate the proposed ensemble, including precision, recall, F1-score, and accuracy. Additionally, a confusion matrix is generated to visualize the distribution of predictions across the \texttt{InteractionActivity} classes, highlighting where the models excel and where they need improvement.

Fig.~\ref{2algotesttrainloss} and Fig.~\ref{onealgotesttrainloss} show the training and validation loss curves for H-SENN, AG-XCaps, and ResNet-GRU, which exhibit a consistent downward trend, indicating effective learning across all architectures. H-SENN and AG-XCaps show relatively stable convergence, while ResNet-GRU steadily improves over subsequent epochs despite starting with a higher initial loss. The gradual decline in validation loss demonstrates the models’ ability to progressively capture and generalize the underlying morphological patterns, reflecting their adaptive learning process. 

\section{Result Analysis and Discussion}

This section presents a comprehensive discussion of the proposed algorithm's results and analysis, detailing its performance using key metrics, including precision, recall, F1 score, and accuracy. We further evaluate the model through an ablation study, providing insights into its complexity by analyzing parameter count and model size. Additionally, the section examines the model's energy efficiency, scalability, robustness, and adaptability, providing a comprehensive view of its operational strengths and potential areas for further improvement. The trailing part of this section comprises the discussion on the Practical Implications of this research study.

The proposed model achieves a precision of \textbf{0.95}, a recall of \textbf{1}, an F1 score of \textbf{0.97}, and an overall accuracy of \textbf{96\%}. These results underscore the model's high reliability in correctly detecting galaxy interactions, indicating an excellent balance between identifying true positives and minimizing false negatives.

\begin{figure}
    \centering
    \includegraphics[width=0.65\linewidth]{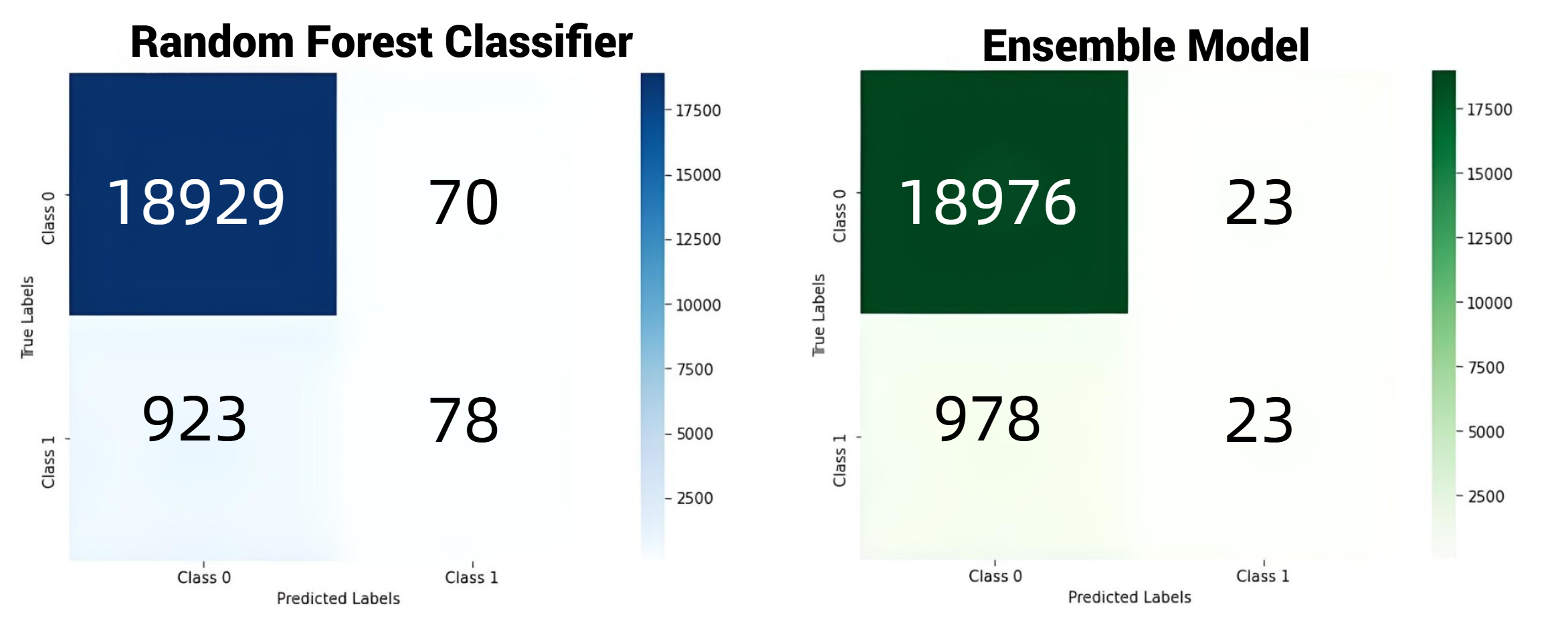}
    \caption{Confusion Matrices Comparing the Random Forest Classifier (left) and the Ensemble Model (right). }
    \label{ConfusionMatrix}
\end{figure}

Comparing the confusion matrices from fig.~\ref{ConfusionMatrix}, the ensemble (23 false positives) significantly reduces the misclassification of normal galaxies compared to the random forest (70 false positives) while maintaining a similar overall performance. It correctly identifies 18,976 normal galaxies and 23 interacting galaxies, whereas the random forest identifies 18,929 normal galaxies and 78 interacting galaxies. Although the ensemble slightly increases false negatives (978 vs. 923), its ability to minimize false positives is crucial when mislabeling normal galaxies as interacting is costly. This trade-off and performance justify the algorithm’s added complexity, demonstrating the benefit of combining multiple models for robust detection of galaxy interactions.

\begin{figure}
    \centering
    \includegraphics[width=0.5\linewidth]{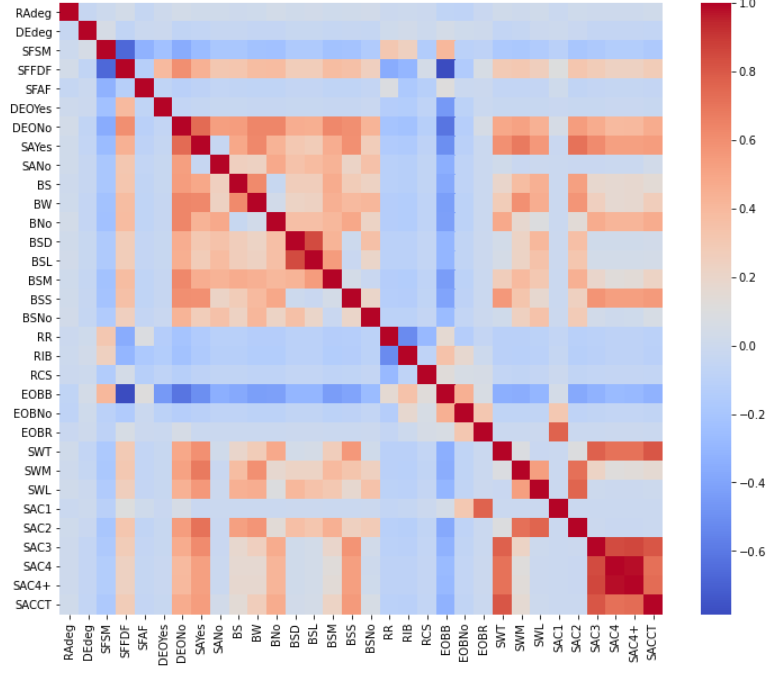}
    \caption{Correlation matrix of the morphological features}
    \label{heatmap}
\end{figure}

Fig.~\ref{heatmap} displays the correlation matrix of morphological features, indicating both positive and negative relationships among variables.

\begin{figure}
    \centering
    \includegraphics[width=0.75\linewidth]{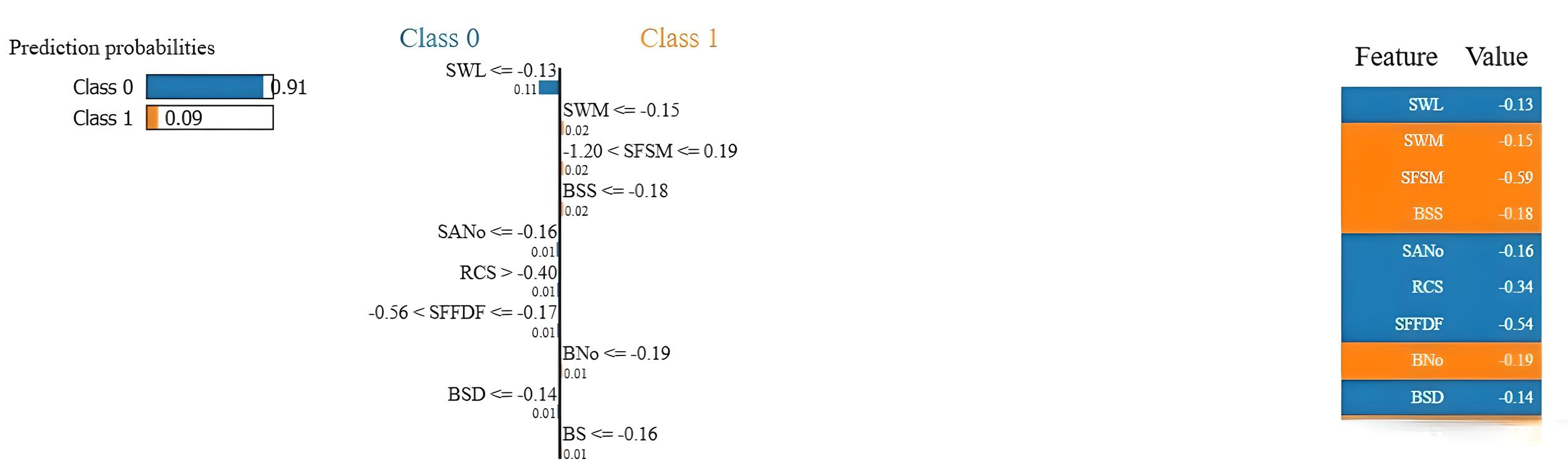}
    \caption{Example LIME Explanation for the Ensemble Model}
    \label{EXLimePred}
\end{figure}

\begin{figure}
    \centering
    \includegraphics[width=0.5\linewidth]{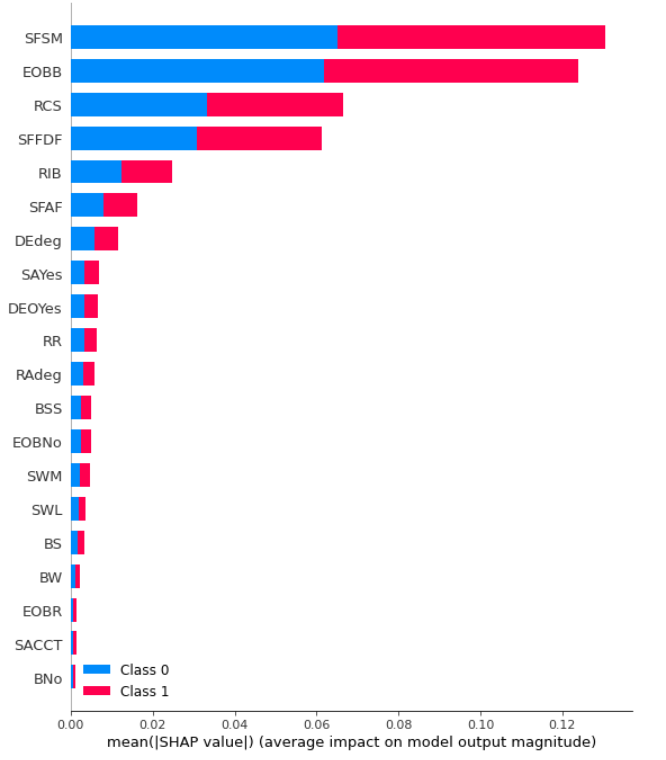}
    \caption{Mean SHAP value bar plot illustrating the contribution of each morphological feature to the classification outcome}
    \label{ShapFeaturePred}
\end{figure}

Fig.~\ref{EXLimePred} visualizes LIME explanations, isolating specific feature contributions to the ensemble's predictions. Complementarily, Fig.~\ref{ShapFeaturePred} displays mean SHAP values, identifying SFSM, EOBB, and RCS as dominant factors. This reliance on features tracking smoothness and asymmetry confirms the model prioritizes physical irregularities—such as tidal distortions—crucial for accurately classifying galaxy interactions.

\subsection{Ablation Study}

\begin{table}[ht]
\centering
\caption{Ablation Study for Individual Models in the Proposed Ensembled Model}
\label{tab:ablation}
\resizebox{0.75\linewidth}{!}{
\begin{tabular}{lcccc}
\toprule
\textbf{Model} & \textbf{Precision} & \textbf{Recall} & \textbf{F1-Score} & \textbf{Accuracy} \\
\midrule
\multicolumn{5}{>{\columncolor[HTML]{FFFFC7}}c}{\textbf{H-SENN}} \\
\midrule
H-SENN\_wo\_Attention   & 0.95 & 1.00 & 0.97 & 92\% \\
H-SENN\_wo\_Residual    & 0.95 & 1.00 & 0.97 & 91\% \\
H-SENN\_wo\_LayerNorm   & 0.95 & 1.00 & 0.97 & 93\% \\
H-SENN\_wo\_Dropout     & 0.95 & 1.00 & 0.97 & 91\% \\
\midrule
\multicolumn{5}{>{\columncolor[HTML]{FFFFC7}}c}{\textbf{ResNet with GRU layers}} \\
\midrule
ResNet\_GRU\_wo\_GRU       & 0.95 & 1.00 & 0.97 & 94\% \\
ResNet\_GRU\_wo\_Residual  & 0.95 & 1.00 & 0.97 & 91\% \\
\midrule
\multicolumn{5}{>{\columncolor[HTML]{FFFFC7}}c}{\textbf{AG-XCaps}} \\
\midrule
AG-XCaps\_wo\_Capsules    & 0.95 & 1.00 & 0.97 & 89\% \\
AG-XCaps\_wo\_Attention   & 0.74 & 0.04 & 0.07 & 71\% \\
AG-XCaps\_wo\_BatchNorm   & 0.95 & 1.00 & 0.97 & 92\% \\
\bottomrule
\end{tabular}
}
\end{table}

As detailed in Table~\ref{tab:ablation}, the ablation study highlights the critical impact of specific architectural components on model performance. For H-SENN, removing attention, residual connections, layer normalization, or dropout yields consistent metrics (precision 0.95, recall 1.00, F1-score 0.97, accuracy 91--93\%), suggesting inherent robustness despite the stability these features provide. Conversely, omitting GRU or residual connections in the ResNet-GRU model reduces accuracy from 94\% to 91\%, underscoring their importance for capturing dependencies. Most notably, AG-XCaps suffers a drastic decline without its attention mechanism, where accuracy plummets to 71\% (precision 0.74, recall 0.04, F1-score 0.07). While removing capsules or batch normalization causes only marginal reductions, these numerical insights confirm that the complex architecture is essential for the effective and robust detection of galaxy interactions.

The ResNet-GRU architecture processes static latent representations from the Adversarial Autoencoder (AAE) by repurposing Gated Recurrent Units (GRUs) to model non-linear dependencies. By treating feature vectors as ordered sequences of morphological descriptors, the model captures higher-order correlations and complex patterns, such as tidal distortions, that dense layers overlook. This leverages sequential processing of static data to robustly augment the ensemble's predictive performance and feature-space analysis.

\subsection{No. of Parameters and Model Size Analysis}

\begin{figure}
    \centering
    \includegraphics[width=0.5\linewidth, height=5cm]{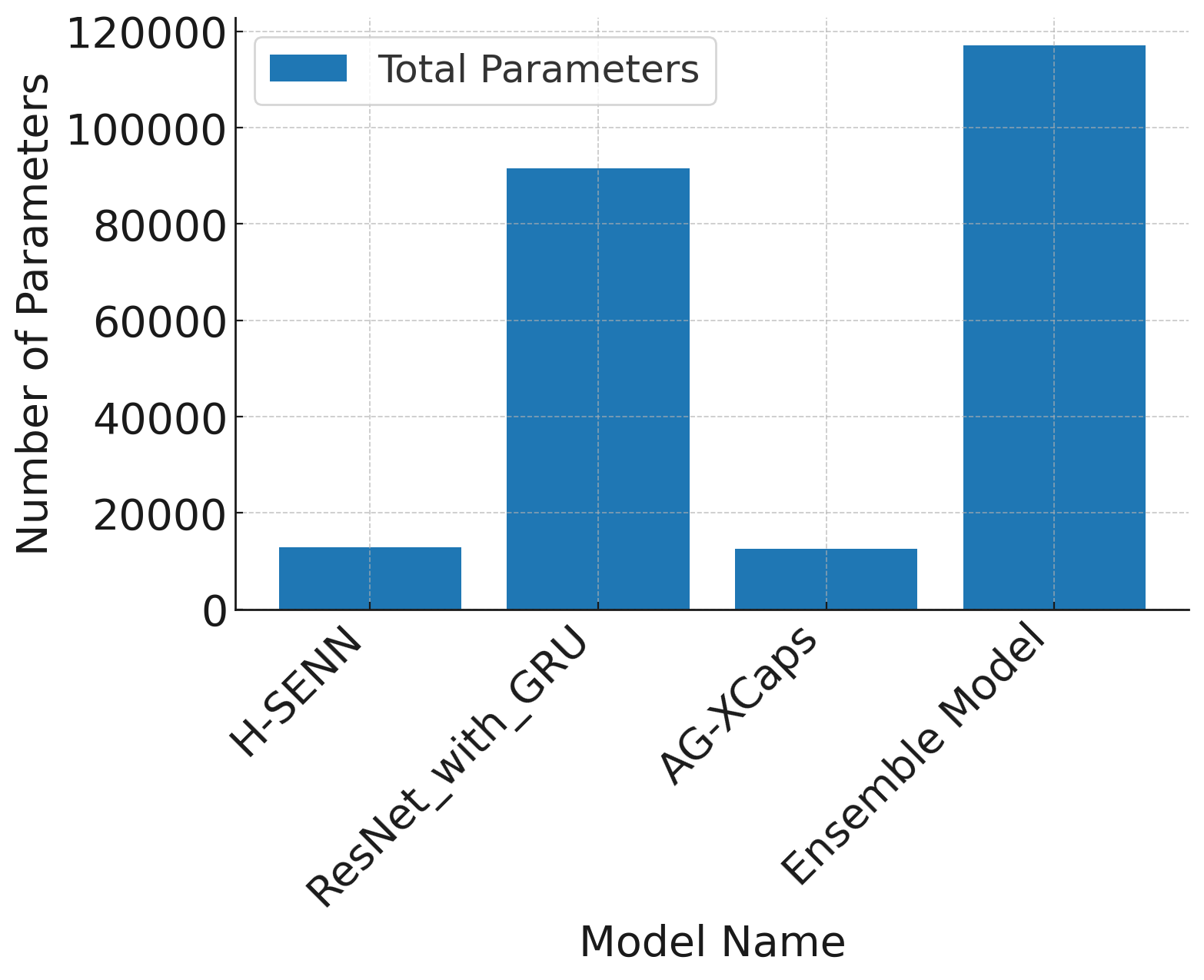}
    \caption{Comparison of the total number of trainable parameters for H-SENN, ResNet\_with\_GRU, AG-XCaps, and the Ensemble Model.}
    \label{parameters}
\end{figure}

\begin{figure}
    \centering
    \includegraphics[width=0.5\linewidth,height=5cm]{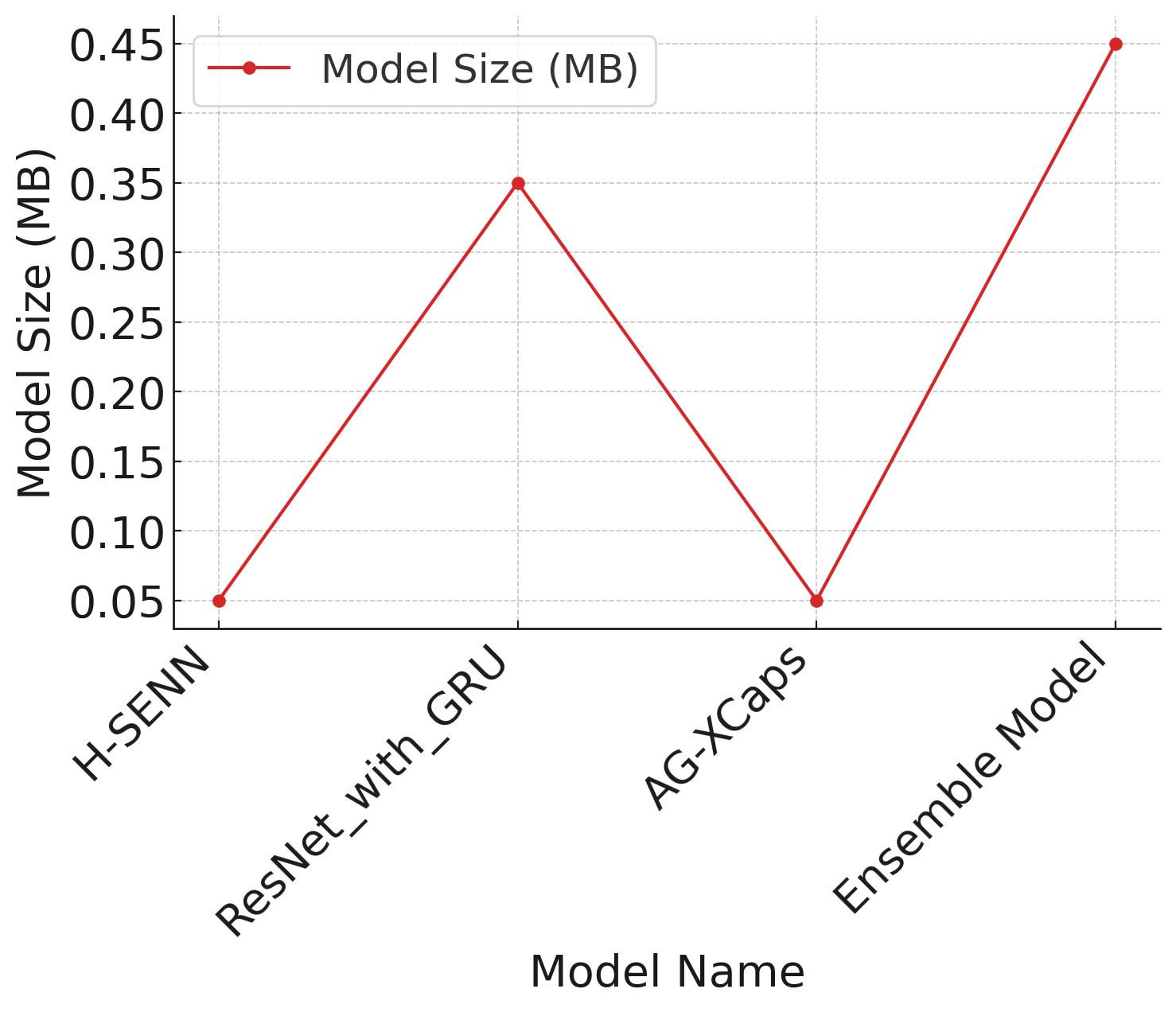}
    \caption{Comparison of Model Size(in MB) for H-SENN, ResNet\_with\_GRU, AG-XCaps, and the Ensemble Model.}
    \label{modelsize}
\end{figure}

Fig.~\ref{parameters} illustrates the number of parameters in the individual models and the ensemble model, while Fig.~\ref{modelsize} provides details on the model sizes of the individual models and the ensemble model. From the two figures above, we infer that the parameter- and model-size comparisons indicate that H-SENN and AG-XCaps require approximately 12,000 parameters (about 0.05 MB). At the same time, ResNet\_with\_GRU has 91,522 parameters (0.35 MB). The ensemble model, which integrates all three architectures, yields 116,998 parameters and a final size of 0.45 MB. Despite this increase, the framework remains lightweight enough for seamless deployment within astronomical infrastructure, offering robust performance without imposing significant computational or storage overhead.

\subsection{Energy, Scalable, Robustness and Adaptive Analysis of the Proposed Model}

This subsection comprehensively evaluates the proposed model across four key dimensions: energy efficiency, scalability, robustness, and adaptability. By examining each aspect, we gain insights into the model's resource intensity, its handling of increasing computational demands, its performance under noisy or adverse conditions, and its generalization across varying data scenarios. Such a multi-faceted analysis ensures a thorough understanding of the model’s performance and highlights areas that may benefit from further optimization.

\begin{figure}
    \centering
    \includegraphics[width=0.5\linewidth]{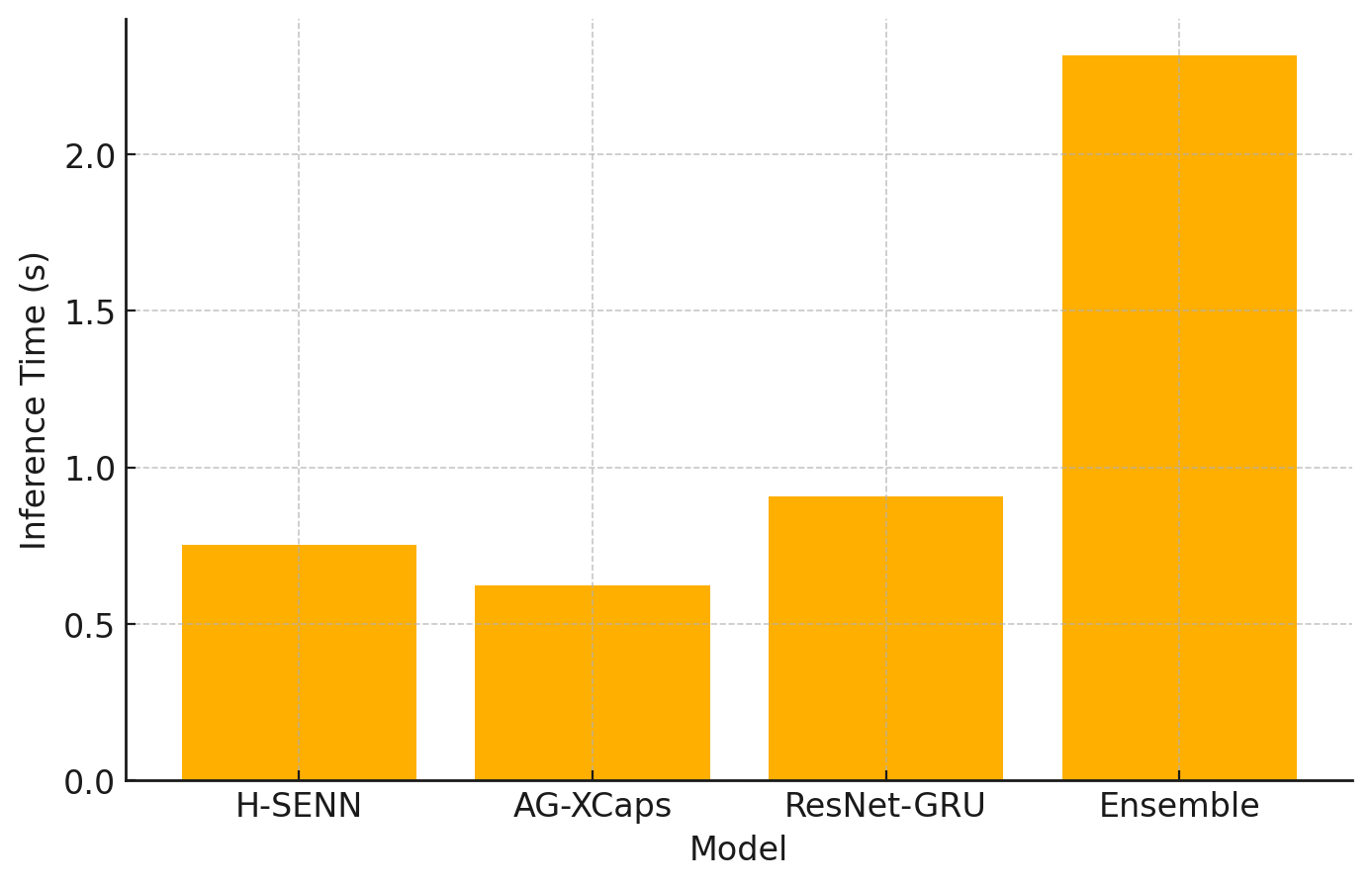}
    \caption{Comparison of inference times (in seconds) for H-SENN, AG-XCaps, ResNet-GRU, and the ensemble model}
    \label{inferencetime}
\end{figure}

\begin{figure}
    \centering
    \includegraphics[width=0.5\linewidth]{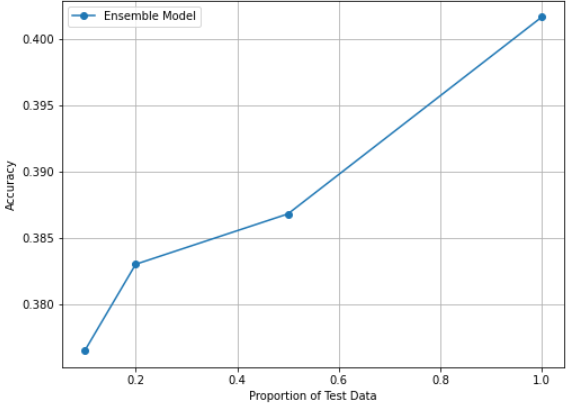}
    \caption{Adaptability Analysis: Accuracy vs. Proportion of Test Data for the Ensemble Model}
    \label{adaptabilityanalysis}
\end{figure}

Fig.~\ref{inferencetime} presents the inference times (in seconds) for H-SENN, AG-XCaps, ResNet-GRU, and the ensemble model, measured by evaluating each model’s prediction on the test set. This indicates energy efficiency, showing that H-SENN and AG-XCaps infer faster than ResNet-GRU. At the same time, the ensemble incurs the highest runtime due to its combined complexity. Fig.~\ref{adaptabilityanalysis} demonstrates the adaptability analysis, showing how the ensemble model’s accuracy improves as the dataset size increases. This confirms the model’s generalizability, indicating that performance improves as larger subsets of the data are used, reflecting enhanced adaptability to diverse input distributions.

\begin{figure}
    \centering
    \includegraphics[width=0.5\linewidth]{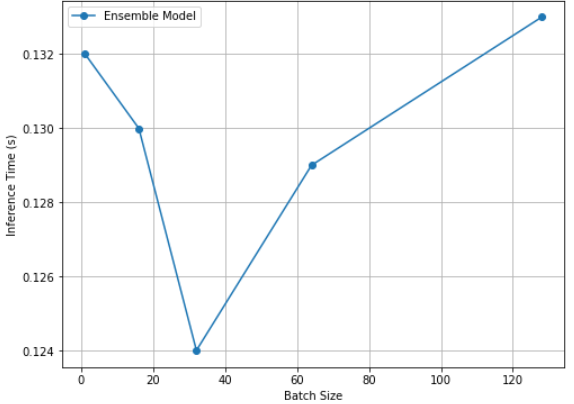}
    \caption{Scalability Analysis: Inference Time vs. Batch Size for the Ensemble Model}
    \label{scalability}
\end{figure}

\begin{figure}
    \centering
    \includegraphics[width=0.5\linewidth]{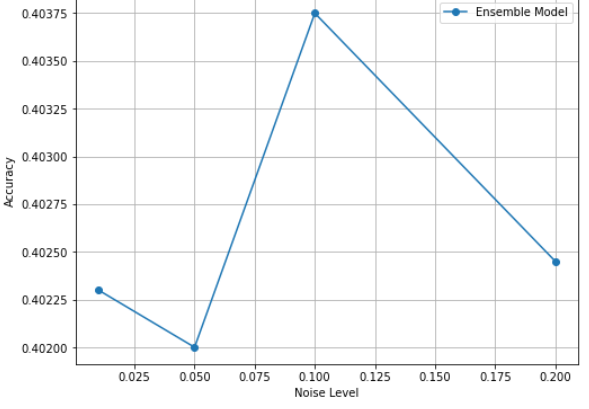}
    \caption{Robustness Analysis: Accuracy of the Ensemble Model under Various Noise Levels}
    \label{robustnessanalysis}
\end{figure}

Fig.~\ref{scalability} illustrates the ensemble model’s inference time across varying batch sizes, demonstrating how simultaneous processing of multiple inputs can reduce latency, thereby reflecting the model’s scalability. However, beyond a certain threshold, inference time increases due to overhead or resource limitations, degrading performance. Complementarily, Fig.~\ref{robustnessanalysis} presents the robustness analysis, in which random noise is incrementally added to the test data to assess variations in accuracy. This evaluation highlights the model’s resilience to imperfect or perturbed inputs, as is typical in real-world conditions, revealing that while performance remains stable at lower noise levels, higher noise intensities lead to a gradual decline in accuracy.

\subsection{Practical Implications of the Research Study}

The proposed explainable galaxy interaction model combines AG-XCaps, H-SNN, and ResNet-GRU with LIME, achieving 0.95 precision, 1.00 recall, 0.97 F1-score, and 96\% accuracy. It enables efficient identification of galaxy interactions in large-scale surveys like DESI, Euclid, and LSST. Its lightweight design (0.45 MB) and scalable inference support real-time deployment on resource-limited infrastructures. LIME enhances transparency by revealing key features (SFSM, EOBB) and aligning results with physical theories and simulations, such as IllustrisTNG. Beyond astronomy, its interpretable, noise-robust, and energy-efficient framework extends to fields like climate modeling and medical diagnostics.

\section{Conclusion}
This study presents a novel attentive neural ensemble technique for explainable galaxy-interaction prediction that integrates AG-XCaps, a Hybrid Self-Attention Neural Network (H-SNN), and ResNet with GRU layers, and is trained on the Galaxy Zoo DESI dataset from the DESI Legacy Imaging Surveys. By leveraging attention mechanisms and Local Interpretable Model-agnostic Explanations (LIME), our model achieves a precision of 0.95, a recall of 1.00, an F1-score of 0.97, and 96\% accuracy, outperforming a Random Forest baseline by reducing false positives (23 vs. 70) while robustly detecting galaxy interactions, as demonstrated by confusion matrix analysis. With a lightweight model size of 0.45 MB, scalable inference, and high noise robustness, our approach provides a transparent and efficient tool for large-scale astronomical surveys, such as Euclid and LSST, significantly advancing understanding of galaxy evolution. Future work could extend this framework to incorporate multimodal data (e.g., spectroscopic and imaging) to improve prediction accuracy, explore real-time processing of survey data streams, and integrate advanced explainability methods (e.g., SHAP) to further refine interpretability for complex astrophysical phenomena.

% \bibliographystyle{IEEEtran}
% \bibliography{ref}

\begin{thebibliography}{99}

\bibitem{walmsley2023galaxy}Walmsley, M., Géron, T., Kruk, S., Scaife, A., Lintott, C., Masters, K., Dawson, J., Dickinson, H., Fortson, L., Garland, I. \& Others Galaxy Zoo DESI: Detailed morphology measurements for 8.7 M galaxies in the DESI Legacy Imaging Surveys. {\em Monthly Notices Of The Royal Astronomical Society}. \textbf{526}, 4768-4786 (2023)
\bibitem{toomre1972galactic}Toomre, A. \& Toomre, J. Galactic bridges and tails. {\em Astrophysical Journal, Vol. 178, Pp. 623-666 (1972)}. \textbf{178} pp. 623-666 (1972)
\bibitem{pearson2019identifying}Pearson, W., Wang, L., Trayford, J., Petrillo, C. \& Van Der Tak, F. Identifying galaxy mergers in observations and simulations with deep learning. {\em Astronomy \& Astrophysics}. \textbf{626} pp. A49 (2019)
\bibitem{nelson2019illustristng}Nelson, D., Springel, V., Pillepich, A., Rodriguez-Gomez, V., Torrey, P., Genel, S., Vogelsberger, M., Pakmor, R., Marinacci, F., Weinberger, R. \& Others The IllustrisTNG simulations: public data release. {\em Computational Astrophysics And Cosmology}. \textbf{6} pp. 1-29 (2019)
\bibitem{ribeiro2016should}Ribeiro, M., Singh, S. \& Guestrin, C. " Why should i trust you?" Explaining the predictions of any classifier. {\em Proceedings Of The 22nd ACM SIGKDD International Conference On Knowledge Discovery And Data Mining}. pp. 1135-1144 (2016)
\bibitem{cao2024galaxy}Cao, J., Xu, T., Deng, Y., Deng, L., Yang, M., Liu, Z. \& Zhou, W. Galaxy morphology classification based on Convolutional vision Transformer (CvT). {\em Astronomy \& Astrophysics}. \textbf{683} pp. A42 (2024)
\bibitem{haslbauer2022high}Haslbauer, M., Banik, I., Kroupa, P., Wittenburg, N. \& Javanmardi, B. The high fraction of thin disk galaxies continues to challenge ACDM cosmology. {\em The Astrophysical Journal}. \textbf{925}, 183 (2022)
\bibitem{laishram2024insights}Laishram, R., Kodama, T., Morishita, T., Faisst, A., Koyama, Y. \& Yamamoto, N. Insights into Galaxy Morphology and Star Formation: Unveiling Filamentary Structures around an Extreme Overdensity at z~ 1.5 Traced by [O ii] Emitters. {\em The Astrophysical Journal Letters}. \textbf{964}, L33 (2024)
\bibitem{urechiatu2024improved}Urechiatu, R. \& Frincu, M. Improved Galaxy Morphology Classification with Convolutional Neural Networks. {\em Universe}. \textbf{10}, 230 (2024)
\bibitem{tsizh5057237galaxy}Tsizh, M., Semenov, V., Tymchyshyn, V. \& Bezguba, V. Galaxy Morphological Classification with Manifold Learning. {\em Available At SSRN 5057237}.
\bibitem{narkedimilli2025predicting}Narkedimilli, S., Raghav, S., Makam, S., Ayitapu, P. \& Others Predicting Stellar Metallicity: A Comparative Analysis of Regression Models for Solar Twin Stars. {\em 2025 IEEE Space, Aerospace And Defence Conference (SPACE)}. pp. 1-6 (2025)


\end{thebibliography}
% \vspace{12pt}

\end{document}